\documentclass[10pt,journal,compsoc]{IEEEtran}

\usepackage{multirow}
\usepackage{booktabs}
\usepackage{amssymb}
\usepackage{subfigure}
\usepackage{enumerate}
\usepackage{svg}
\usepackage{cite}

\usepackage{algorithm}  
\usepackage{algorithmic}

\usepackage{graphicx}
\ifCLASSINFOpdf

\else

\fi

\usepackage{amsmath}
\usepackage{algorithmic}
\usepackage{array}

\usepackage{url}

\usepackage{xcolor}

\begin{document}

\title{ A Data-driven Approach for Constructing Multilayer Network-based Service Ecosystem Models}

\author{
\IEEEauthorblockN{
Mingyi Liu\IEEEauthorrefmark{1},
Zhiying Tu\IEEEauthorrefmark{1},
Xiaofei Xu\IEEEauthorrefmark{1}, 
Zhongjie Wang\IEEEauthorrefmark{1}}\\
\IEEEauthorblockA{
\IEEEauthorrefmark{1}School of Computer Science and Technology,
Harbin Institute of Technology, Harbin, China}
\thanks{Manuscript received December XX, XXXX; revised August XX, XXXX. 
Corresponding author: Z. Wang (email: rainy@hit.edu.cn).}}

\markboth{Journal of \LaTeX\ Class Files,~Vol.~XX, No.~XX, August~XXXX}%
{Shell \MakeLowercase{\textit{et al.}}: Bare Demo of IEEEtran.cls for Computer Society Journals}

\IEEEtitleabstractindextext{
\begin{abstract}

Services are flourishing drastically both on the Internet and in the real world. Additionally, services have become much more interconnected to facilitate transboundary business collaboration to create and deliver distinct new values to customers. Various service ecosystems have become a focus in both research and practice. However, due to the lack of widely recognized service ecosystem models and sufficient data for constructing such models, existing studies on service ecosystems are limited to very narrow scope and cannot effectively guide the design, optimization, and evolution of service ecosystems. We propose a Multilayer network-based Service Ecosystem Model, which covers a variety of service-related elements, including stakeholders, channels, functional and nonfunctional features, and domains, and especially, structural and evolutionary relations between them. "Events" are introduced to describe the triggers of service ecosystem evolution. We propose a data-driven approach for constructing MSEM from public media news and external data sources. Qualitative comparison with state-of-the-art models shows that MSEM has a higher coverage degree of fine-grained elements/relations in service ecosystems and richer semantics for higher interpretability. Experiments conducted on real news corpora show that compared with other approaches, our approach can construct large-scale models for real-world service ecosystems with lower cost and higher efficiency.
\end{abstract}

}

\maketitle

\IEEEdisplaynontitleabstractindextext

\IEEEpeerreviewmaketitle

\section{Introduction}\label{sec:introduction}
The cloud, the Internet of Things (IoT), and various virtualization technologies have sharply increased the number of available services.
Services have become increasingly 
interconnected to facilitate transboundary collaboration for creating and delivering unique new value to customers. Many researchers have focused on this new phenomenon and invented various new terms for it, such as ``Internet of Services''\cite{schroth2007web}, ``Big Services''\cite{xu2015big}, and 
``Crossover Services''\cite{wu2015modern}. All of these terms are used to describe complicated service ecosystem phenomena but with different theoretical focuses.

 \begin{figure*}[!htbp]
     \centering
     \includegraphics[width=0.9\textwidth]{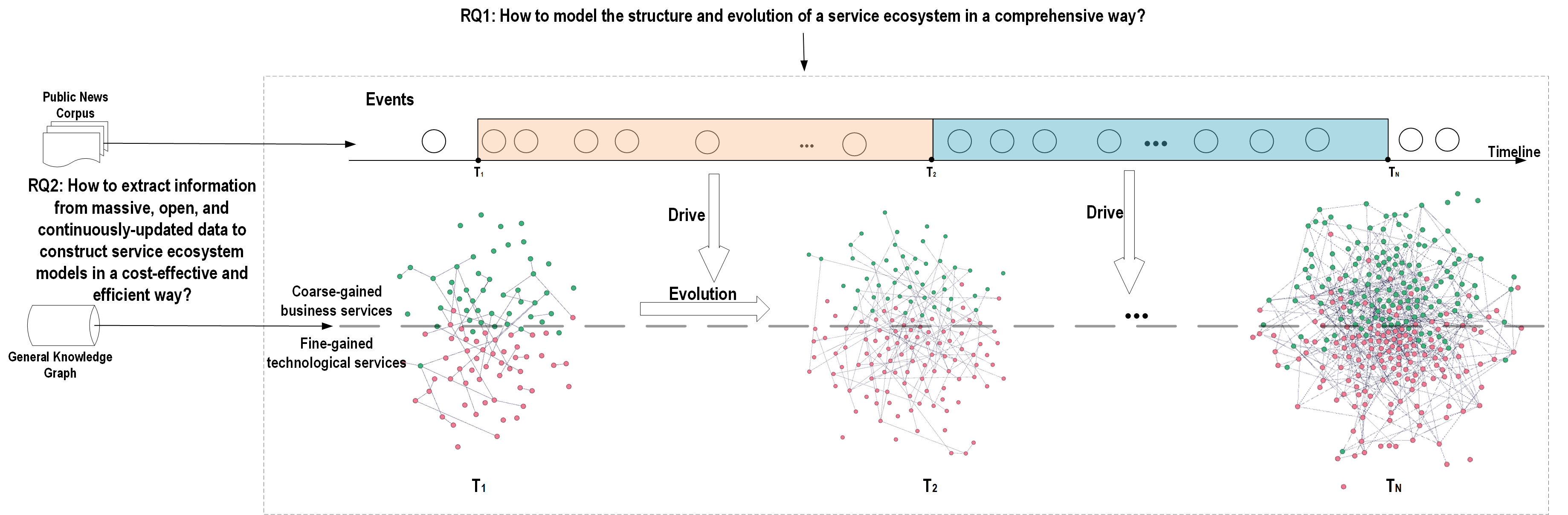}
     \caption{Structure and evolution of a service ecosystem}
     \label{fig:abstract}
 \end{figure*} 
As a new phenomenon that has quickly dominated many modern service industries but lacks sufficient theoretical foundations, service ecosystems have drawn considerable attention from the academic community in recent years. Some researchers have focused on the network attributes of service ecosystems\cite{adeleye2018constructing,zhou2019clustering}.

Other researchers have applied service ecosystem concepts to traditional service computing problems such as service discovery, service composition, and service recommendation \cite{huang2014recommendation, wang2018dkem, zhang2015dynamic}.

In addition to the research outcomes mentioned above, we believe that the service ecosystem concept can also contribute to a variety of business-level problems, including the following: (1) What does the structure of a service ecosystem look such as to precisely delineate business collaborations among organizations, and how are such business collaborations enabled under the support of technological service collaboration? (2) How does a service ecosystem come into being and evolve over time? (3) Why does a service ecosystem keep evolving? Answering these problems can bring significant benefits to both service providers and market regulators who are involved in the creation and evolution of service ecosystems. A service provider can learn its competitors' collaboration strategies and popular/subtle evolution trends of service collaboration from a global point of view so that it can adjust its innovation strategy and business collaboration strategy to identify those fleeting innovation opportunities, thereby enhancing its competitiveness in the service market. For market regulators, understanding the evolution of service ecosystems can facilitate the healthy and sustainable development of the service market by applying appropriate incentives, such as issuing guiding policies and laws. An elaborate model for describing the structure and evolution of service ecosystems is highly required to address the abovementioned challenges.

The lack of a comprehensive service ecosystem model and the inadequacy of real-world data for constructing such models limit the depth of service ecosystem research. Deficiencies in the current research are twofold:

(1) Most of the existing models are partial and only cover a single perspective of service ecosystems. There is a lack of a comprehensive view of multilevel service ecosystems. In our opinion, a service ecosystem is a complex sociotechnological system that is composed of a variety of entities, such as business domains, organizations, service APIs, physical service facilities, and event their features. A good model should be able to delineate the characteristics of a service ecosystem at both the business and technology levels, in both digital and real worlds, and in both functional and nonfunctional perspectives. Only focusing on one of them would neglect a large amount of complementary relation information among different perspectives.

(2) Existing service ecosystem modeling approaches are mostly based on a small-scale dataset, such as web APIs publicized on the ProgrammableWeb\footnote{\url{https://www.programmableweb.com/}}.

This small and technique-related dataset not only damages the credibility of the constructed models but also hinders the usage of those popular and powerful deep learning methods in service ecosystem analytics for deeper insights. In our opinion, service ecosystem modeling should not be based on a single data source but should make full use of multiple types of open data, even though those data are unstructured or unannotated. Another neglected issue is the timeliness of data: a long time lag between actually occurring changes and lagging data collection prevents the constructed models from fully exhibiting the latest states of real-world continuously evolving service ecosystems. As a consequence, service providers and market regulators cannot timely perceive those trivial but possibly influential changes from the models. The last deficiency is the lack of adequate semantics in datasets, which makes the constructed service ecosystem models difficult to interpret. 

As Fig. \ref{fig:abstract} shows, the service ecosystem can be modeled as a kind of complex network. Therefore, if we present it over time, then we might determine its evolution caused by timely ``external events". To do this, we should answer the following two research questions:

\begin{enumerate}
\item[\textbf{RQ1}]  How can the structure and evolution of a service ecosystem be comprehensively modeled?
    \item[\textbf{RQ2}] How can information be extracted from massive, open, and continuously updated data to construct service ecosystem models in a cost-effective and efficient way?
\end{enumerate}

Following the philosophy discussed above, this paper proposes a novel \underline{M}ultilayer network-based \underline{S}ervice \underline{E}cosystem \underline{M}odel (MSEM) and a data-driven approach for constructing MSEM from massive public media news and external data sources. Knowledge graphs (KGs), natural language processing (NLP) techniques and joint learning methods are jointly utilized for the construction of MSEM. The main contributions and innovations are as follows:
\begin{itemize}
    \item  Our MSEM comprehensively covers essential elements of stakeholders, services delivery channels,  service  functional  features  and  nonfunctional  features, and business domains, and the interconnections between them. This ensures the completeness of service ecosystem models. 
    \item Events are imported into MSEM as triggers of service ecosystem evolution. This ensures that the model has the capacity to actively perceive the evolution of service ecosystems.
    \item Our approach not only leverages structured data from knowledge graphs but also extracts rich semantics from massive open unstructured data. This ensures that MSEM can be used for more valuable analytics and reasoning at the business level.
    \item Metrics including \textit{cost}, \textit{coverage}, \textit{interpretablity}, \textit{semantic integrity} and \textit{semantic accuracy} are used to qualitatively evaluate the usefulness and usability of MSEM, and comparisons have shown that our model outperforms existing models. Comparative experiments on real data reveal that our model construction method is with lower cost and higher efficiency.
\end{itemize}

The remainder of this paper is organized as follows. Section \ref{sec:related_work} introduces related work. Section \ref{sec:model} defines the metamodel of MSEM in detail. Section \ref{sec:construction} gives the process and steps of constructing MSEM from external data sources. Section \ref{sec:compare} makes a qualitative comparison between the MSEM and other existing service ecosystem models. Section \ref{sec:eval} evaluates the performance of the data-driven MSEM construction approach. Section \ref{sec:application} gives two real-world examples to illustrate potential application scenarios of MSEM. and the last section is the conclusion

\section{Related Works}\label{sec:related_work}
\subsection{Natural Ecosystem}
The term ``ecosystem'' was first used in the field of biology\cite{willis1997ecosystem}. It was originally defined as ``\textit{a community of living organisms with air, water and other resources}''. Later, its connotation was extended to ``\textit{a community of living organisms in conjunction with the nonliving components of their environment, interacting as a system}'' \cite{molles2015ecology, chapin2011principles, smith2012elements}.

Kaufman\cite{kauffman1996home} argued that the formation of an ecosystem is influenced by some attractors (those no-biological components) taht are regarded as ``resources'' of the ecosystem. The quantity and quality of available resources push the formation of stable communities of species. An ecosystem is not static, and according to Rosen\cite{rosen2000world}, ``\textit{ecosystems are dynamic, constantly remaking themselves, reacting to natural disturbances and to the competition among and between species}''. This indicates that an ecosystem is an evolving system, and it is capable of adapting to changes from outside or inside of it by reorganizing communities and relations among species.

\subsection{Digital Ecosystem}
Inspired by natural ecosystems, many studies have applied ecological theory to information systems, known as digital ecosystems. A digital ecosystem is a distributed, adaptive, open sociotechnical system with properties such as self-organization, scalability, and sustainability. Digital ecosystem models are built based on the knowledge of natural ecosystems, especially on aspects related to competition and collaboration among diverse entities\cite{briscoe2006digital, zhu2015digital, dini2005digital}. Unlike natural ecosystems, digital ecosystem research covers a wide variety of domains.

For example, Mitchell et al \cite{mitchell2007opensocial} proposed a social ecosystem framework called \textit{OpenSocial} to align online digital world with physical world. This framework enables different social networks to link with each other and to self-organize into a social ecosystem under the policy guidance of individuals and organizations; Hazenberg et al \cite{hazenberg2016role} built a social enterprise ecosystem (SE-ecosystem) to explore the development of stakeholder and institutional networks across Europe; Broring et al \cite{broring2017enabling} developed an Internet of Things ecosystem (IoT-ecosystem) to eliminate interoperability barriers among different IoT systems. Peltoniemi et al \cite{peltoniemi2004business} explained the concept of business ecosystems and used it to analyze and explain continuously changing business environment; Evans et al\cite{evans2016revealing} explored an approach of developing business strategies for enterprises by a visual study on API ecosystems (API-ecosystem).

The digital ecosystem topic has been very hot in academic communities of both business and technology because researchers have fully realized the power of ecosystem theory in delineating large and complicated sociotechnological or manufactured systems.

\subsection{Service Ecosystem}
A service ecosystem is a special type of digital ecosystem, but there has not yet been a widely recognized definition. Because services demonstrate the \textit{business-technology duality}, the corresponding service ecosystems should be considered as the combination of business ecosystems and digital ecosystems. A service ecosystem is comprised of (a) entities acting in domain-specific roles (e.g. providers and consumers), (b) services available for business collaboration and value cocreation, and (c) infrastructure for realizing service engineering, delivery and governance\cite{ruokolainen2012framework}. Researchers have focused on the architecture, models, and creation and evolution mechanisms of service ecosystems. 

For example, Wu et al\cite{wu2015modern} proposed a modeling framework for crossover services in which services from different domains collaborate together to create new values that a single service cannot provide; Xu et al\cite{xu2015big} modeled a service ecosystem as a multilayer network in which services are aggregated from the bottom up layer by layer, and there form a set of frequently used service chains and service hyperchains that are defined as ``transboundary service patterns''. Studies that apply the service ecosystem concept to traditional service computing problems, such as service discovery\cite{adeleye2018constructing}, service selection\cite{wang2018dkem},  and service recommendation\cite{huang2014recommendation}, have appeared frequently at service computing conferences in recent years.

\section{Multilayer Network-based Service Ecosystem Model (MSEM)} \label{sec:model}

Inspired by the natural ecosystem definition, in this paper, we define a service ecosystem as a community of stakeholders in conjunction with the services they offer or use. In this community, stakeholders interact through service offerings and service consumption, and they form stable or occasional service interconnections at both the business and technological levels. This ecosystem continues to evolve by responding to stimuli from inside or outside.

\begin{figure*}
    \centering
    \includegraphics[width=0.9\textwidth]{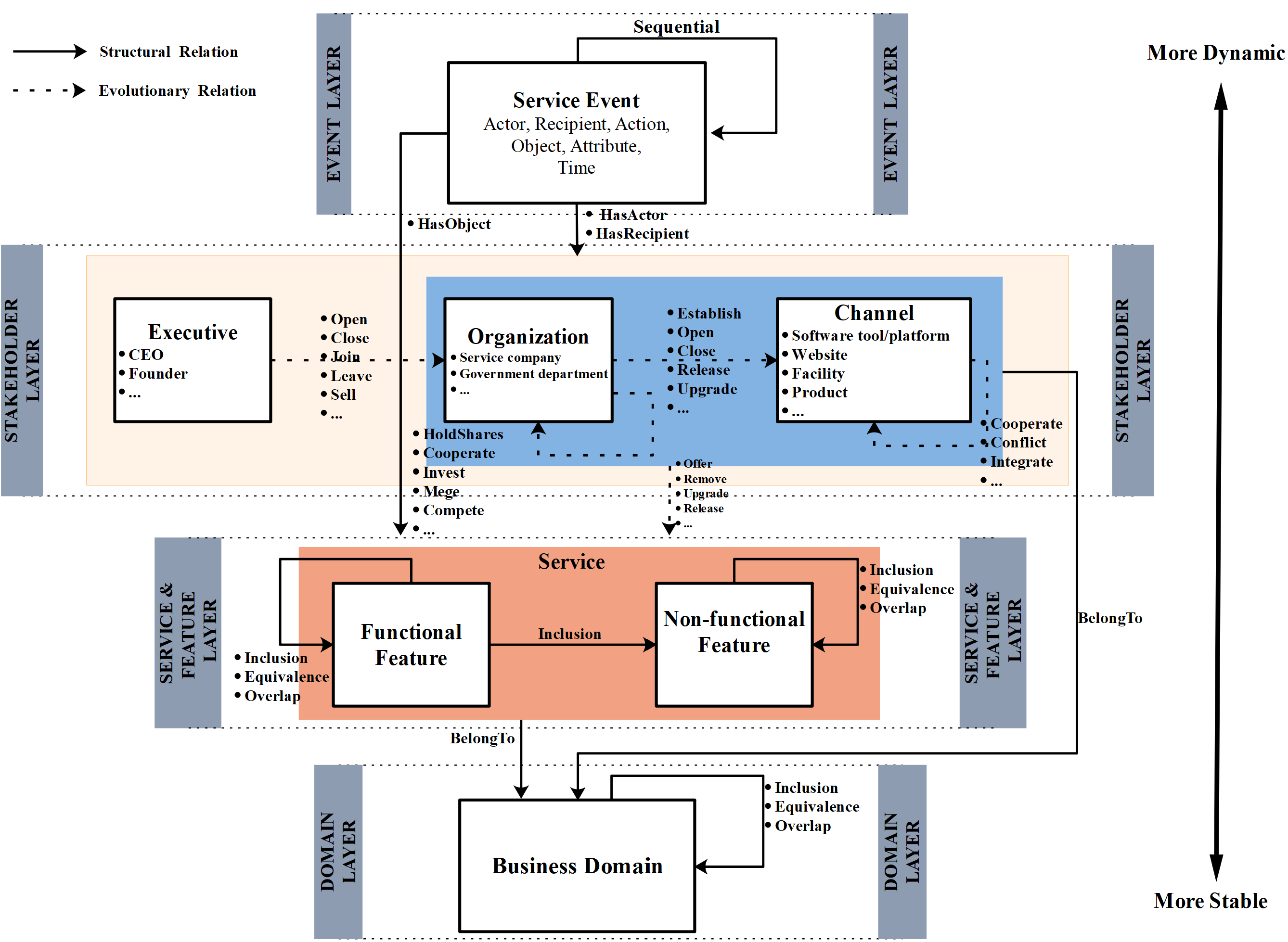}
    \caption{Illustration of the multilayer network-based service ecosystem model (MSEM)}
    \label{fig:overview}
\end{figure*}

Based on the definition discussed above, we propose a multilayer network based service ecosystem model (MSEM) that is composed of four layers: the \textbf{event layer}, \textbf{stakeholder layer}, \textbf{service \& feature layer}, and \textbf{domain layer}. Elements in each layer form a network structure, and interlayer relations connect networks of four layers into a holistic layered network. Since this model is evolution-oriented, in addition to traditional \textit{structural} and \textit{semantic} relations, we import \textit{evolutionary} relations to depict the actions that may occur between two elements. Fig. \ref{fig:overview} shows an overview of this model, in which solid arrows are structural relations, while dotted arrows are evolutionary relations.

\subsection{Event Layer}\label{sec:event_layer}

Events are \textit{things that happen} \cite{rospocher2016building}. A service-related event is an action that is initiatively taken by a stakeholder, acts on other stakeholders or services, and possibly results in the changes of elements and evolutionary relations in a service ecosystem. 

In this paper, an event consists of six components:
\begin{itemize}
    \item \textbf{Actor}: a stakeholder who initiates the event.
    \item \textbf{Action}: a behavior that is conducted by the \textbf{Actor} of the event (usually a \textit{verb}).
    \item \textbf{Recipient}: a stakeholder who is passively acting in the event.
    \item \textbf{Object}: a purpose or effect of the event, usually functional or nonfunctional features of a service. 
    \item \textbf{Attribute}: additional information of the event (may be numeric or descriptive).
    \item \textbf{Time}: the time when the event occurs.
\end{itemize}

There are two types of relations between events, i.e., \textbf{sequential} and \textbf{causal} \cite{li2017eeg}. Our empirical study shows that in the news corpus, causality is rarely explicitly expressed in syntaxes such as \textit{X because of Y} or \textit{Y causes X} but is usually expressed implicitly. Actually, \textbf{causal} is a subtype of \textbf{sequential}; thus, in the network model of this layer, only \textbf{sequential} is kept as a structural relation.

Due to the prosperity of the media industry, almost any significant, subtle or not obvious actions that have occurred in the real world can be reported and publicized in a timely manner on a variety of social media. We extract events from the news corpus published on social media (to be discussed in Section \ref{sec:extraction_news}) and obtain time-series events to form the model in this layer (see the upper part of Fig. \ref{fig:abstract}).

\subsection{Stakeholder Layer}
The network in this layer is composed of coarse-grained business services in a service ecosystem. Nodes in this layer are stakeholders that are involved in a service ecosystem and have business collaboration with each other. A stakeholder can take active actions or passively respond to actions taken by others. There are three types of stakeholders:

\begin{itemize}
    \item \textbf{Organization}: including service companies (e.g., \textit{Tencent}, \textit{Alibaba}) and government departments (regulators, e.g., \textit{State Post Bureau of China}).
    \item \textbf{Channel}: being a carrier of service delivery, a channel is developed/operated by an organization and individuals or other organizations use the channel to access specific service features to meet their special demands. Channels may be software services (e.g., a mobile app \textit{WeChat}, a web API \textit{Skyscanner Flights API}), tangible products (e.g., \textit{iPhone 11}), or physical facilities in the real world (e.g., \textit{Amazon Go} stores).
    \item \textbf{Executive}: Key persons who play important roles in an organization and influence what actions should be taken by the organization, such as founders or CEOs of a company.
\end{itemize}

Relations between two stakeholders are \textit{evolutionary}, i.e., there is an attached timestamp indicating when a relation comes into being. In the specification of MSEM, we do not limit specific types of evolutionary relations but only give some frequently appearing referential types (listed on the arrows of Stakeholder Layer of Fig. \ref{fig:overview}), and more types of emerging relations may be extended dynamically during model construction. This enhances the extensibility of the model.

All the evolutionary relations can be derived from events and are thus extensible as long as a new type of evolutionary relation identified from news corpus has an interpretable meaning and an event timestamp. Generally, an identified event would trigger the addition/removal of one or multiple stakeholders, services, and features, and
the addition/removal of one or multiple evolutionary relations between stakeholders. An evolutionary relation is thus represented by a quintuple
(\textit{source}, \textit{destination}, \textit{relation}, \textit{timestamp}, \textit{additionalAttributes}), where all these components are mapped from the corresponding event components. Details of such mapping are discussed in Section \ref{sec:phase3}.

\subsection{Service \& Feature Layer}
The network in this layer is composed of fine-grained technological services in a service ecosystem. There are three types of nodes, i.e., services, functional features, and nonfunctional features. A service is composed of a set of functional and nonfunctional features. A functional feature represents a functional entity that can meet a specific user demand (e.g., \textit{online payment}, \textit{instant messaging}), and a nonfunctional feature represents a cross-cutting concern on functional features, such as \textit{intellectual property rights}, \textit{service reliability}, and \textit{customer satisfaction}). All three types of nodes can be the \textbf{object} of an \textit{event}.

Note that in MSEM, there is a clear distinction between a channel and a service. In the services computing community, researchers usually regard software entities such as web services, web APIs, and mobile apps as ``services'' that offer specific functionalities via well-defined interfaces. In our model, we separate \textit{logical} services from \textit{physical} channels to make the model clearer and more precise. In other words, a channel is a physical entity through which logical services along with functional and nonfunctional features are offered to users. Logical services may be offered through multiple different channels (e.g., \textit{flight search} can be accessed via the \textit{Skyscanner website}, \textit{Skyscanner mobile app}, or \textit{Skyscanner flights API}), and one channel can offer multiple logical services (e.g., \textit{WeChat} offers services including \textit{social networking}, \textit{online payment}, and \textit{various public services}).

Relations between these nodes are structural and semantic-based, i.e., they are identified from the conceptual level and do not change with time. There are three types of relations: \textbf{equivalence}, \textbf{inclusion}, and \textbf{overlap}.
For example, \textit{instant messaging} \textbf{includes} \textit{video calls}, \textit{blockchain currency} and \textit{supply chain traceability} \textbf{overlap} in \textit{blockchain service}, and \textit{WeChat payment} and \textit{Alipay payment} are both for \textit{online payment}, so they are \textbf{equivalent}.

\subsection{Domain Layer}
The network in this layer is composed of \textit{domain} nodes, such as healthcare, transportation, retailing, logistics, social, finance, and so on, and relations between domains. Similar to the service\&feature layer, relations between domains are structural and are classified into \textit{equivalence}, \textit{inclusion}, and \textit{overlap}. They are less likely to change.

\subsection{Relations Between Layers}
There are cross-layer relations between elements of different layers.

An event is composed of event components, each of which can be mapped to a corresponding entity in the stakeholder layer or the service \& feature layer. \textbf{HasX} is used to represent an interrelation between these layers, where \textbf{X} can be one of the three event components: \textbf{actor}, \textbf{recipient}, and \textbf{object}. \textbf{HasX} is a structural relation that does not change over time.

Any entities in the stakeholder layer and service \& feature layer tend to be associated with one or more domains. \textbf{BelongTo} is used to connect the stakeholder and service \& feature layer with the domain layer. It is also a structural relation. For example, \textit{WeChat} and \textit{Facebook} belong to the \textit{social} domain, and since \textit{WeChat} also provides the service \textit{WeChat payment}, it also belongs to the \textit{finance} domain.

The stakeholder layer and the service \& feature layer are the kernel layers of the model. Relations between them are similar to the \textit{demand for resources} in natural ecosystems and are evolutionary. Stakeholders \textit{offer} functional features to meet user requirements and \textit{upgrade} nonfunctional features to enhance user experiences. Stakeholders may also \textit{release} some new functional features or \textit{close} some unpopular functional features based on market demands. For example, \textit{Google} offers a functional feature \textit{search engine} and can upgrade its nonfunctional feature \textit{search speed} by allocating more computing resources; on April 2, 2019, \textit{Google} closed its troubled \textit{Google+} channel and the corresponding \textit{social network service}.

\section{A Data-driven Approach for MSEM Construction}\label{sec:construction}
\begin{figure*}[!thbp]
    \centering
    \includegraphics[width=0.9\textwidth]{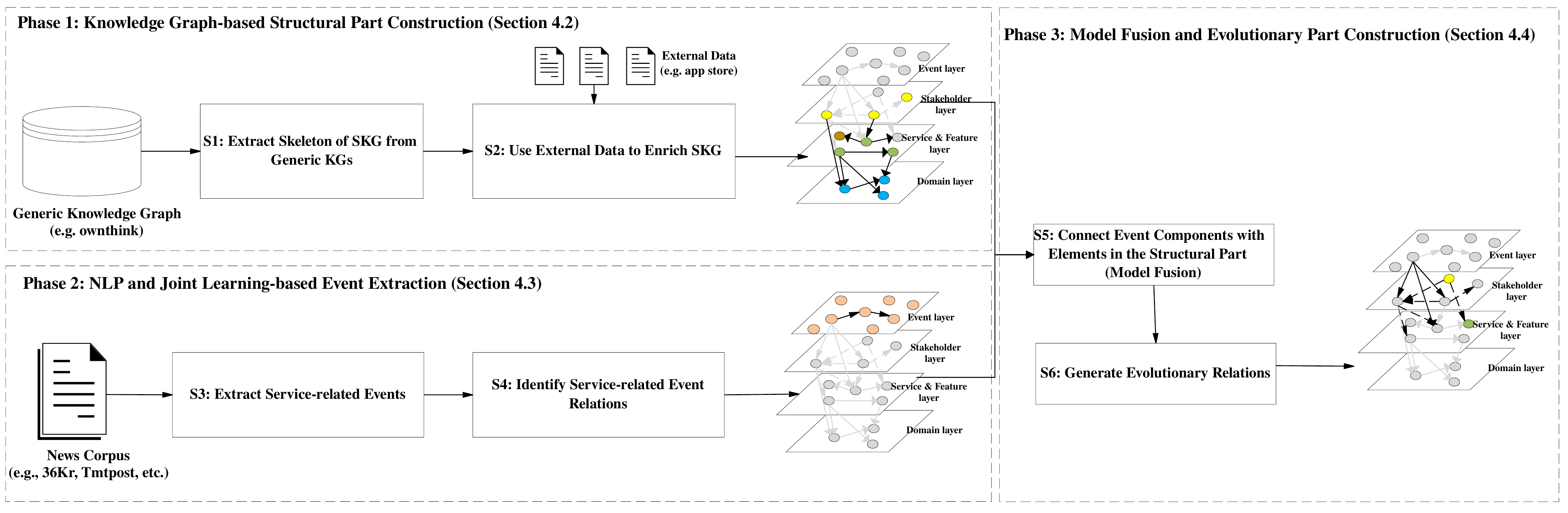}
    \caption{A data-driven approach for constructing a multilayer network-based service ecosystem model (MSEM)} 
    \label{fig:construction}
\end{figure*}

\subsection{MSEM Construction Process}

From Section \ref{sec:model}, we can see that there are two types of elements in MSEM: (1) stable elements including \textit{organizations}, \textit{channels}, \textit{executives}, \textit{services}, \textit{functional features}, \textit{nonfunctional features}, and \textit{domains}; and (2) \textit{events} that occur dynamically/irregularly and trigger the evolution of other elements and relations. There are two types of relations: (1) \textit{structural relations} that are semantically-based and usually remain stable and (2) \textit{evolutionary relations} that are dynamic and triggered by events. Therefore, we divide an MSEM into two parts:

\begin{itemize}
    \item Structural part: including stable elements and structural relations between them;
    \item Evolutionary part: including events and evolutionary relations that events trigger.
\end{itemize}

We follow a straightforward \textit{data-driven} philosophy for MSEM construction: the structural part is constructed with the help of a generic knowledge graph and external data sources, and the evolutionary part is constructed from a publicized news corpus. Fig. \ref{fig:construction} shows a schematic overview of this process, which is decomposed into three phases:

\begin{enumerate}[{Phase} 1]

\item Knowledge graph-based structural part construction: to extract entities in the \textit{stakeholder layer},
\textit{service \& feature layer}, and \textit{domain layer} from generic KG and external public data sources, as well as to identify structural relations between these entities.
\item NLP and joint learning-based event extraction:
to identify events from the public news corpus, including six components of each event and relations between events. NLP and joint learning approaches are employed in this phase.
\item Model fusion and evolutionary part construction: to fuse the results of Phase 1 and Phase 2 by connecting event components with elements in the structural part (i.e., to identify those cross-layer structural relations in Fig. \ref{fig:overview}) and to identify evolutionary relations between elements in \textit{the stakeholder layer} and between elements in \textit{stakeholder layer} and in \textit{service \& feature layer}.
\end{enumerate}

The reminder of this section introduces the technical details of each phase.

\subsection{Phase 1: Knowledge Graph-based Structural Part Construction}\label{sec:knowledge}
The structural part constitutes the skeleton of MSEM. Compared with evolutionary elements such as events and evolutionary relations, elements in the structural part are more stable, and the relations between them are semantically based. The construction of the structural part is based on a knowledge graph.

A KG is a collection of interlinked descriptions of entities
and relationships between entities are usually tagged with types that provide information about the nature of the relationship.
Google, Facebook, and many other corporations have devoted many resources to building large-scale KGs for their business, and there have been many open source KGs publicized on the Internet, such as DBpedia and Freebase.

Since KGs are usually constructed based on the rich information available on the Internet, we believe that most stable elements in MSEM and structural relations between them should have been included in existing large-scale KGs. This is the reason why we used KGs in this phase.

To the best of our knowledge, there are no available open source service-domain KGs; thus, we have to switch to generic KGs. However, (1) generic KGs contain a large number of entities unrelated to services, and these entities are of no use to MSEM; (2) generic KGs cannot fully cover all the service-related entities and their relations existing in the real world, which hinders the scale of the MSEM. Considering the two issues, we construct a \underline{S}ervice-domain \underline{K}nowledge \underline{G}raph (SKG) based on the refinement of a generic KG named \textit{ownthink}\footnote{https://www.ownthink.com/knowledge.html} and then extend it by external data sources.

First, a set of rules in the form of regular expressions are generated manually by domain experts, and they are used to remove those service-unrelated entities from the generic KG and classify the remaining service-related entities into concrete types of stable elements in MSEM. This is step S1 in Fig. \ref{fig:construction}. The rules we used can be found in GitHub\footnote{https://github.com/icecity96/TSC2019appendix}.

Next, to let SKG cover more service-related entities, we collect additional \textit{organizations} and \textit{executives} information from \textit{PEdaily}, which contains the above $86,500$ companies worldwide, and most of them are service-related companies. Apps are a typical channel, and now, an increasing number of organizations provide their services through apps.
Therefore, we collect \textit{organizations}, \textit{channels}, \textit{services} and \textit{features} information from the \textit{MI App store}.
Note that it is not limited to the two external data sources, but more data sources can be utilized. A complete data source list and crawlers we used can be found on GitHub\footnote{https://github.com/icecity96/serviceKnowledgeSpider}.

Finally, we use \textit{shpy}\footnote{https://github.com/Beim/shpy}, a tool we developed to manage heterogeneous external data resources and fuse them with the refined KG through an entity alignment method called RiMOM\cite{li2008rimom}.
This is step S2 in Fig. \ref{fig:construction}. Details of the SKG we built can be found in Section \ref{sec:dataset}.

\subsection{Phase 2: NLP and Joint Learning-based Event Extraction}\label{sec:extraction_news}

As shown in Fig. \ref{fig:construction}, there are two tasks in this phase:
\begin{itemize}
    \item \textbf{Event Extraction (S3)}: to extract service-related events from news corpus. By treating six components of service events as named entities, this can be viewed as a \textit{named entity recognition} (\textit{NER}) task or \textit{semantic role labeling} (\textit{SRL}) task in NLP.
    \item \textbf{Event Relation Identification (S4)}: to classify the relation between two given events into given types, i.e., \textit{unrelated}, \textit{sequential}, and \textit{reverse sequential}.
\end{itemize}

\begin{figure}[hbtp]
 \centering
 \includegraphics[width=0.9\linewidth]{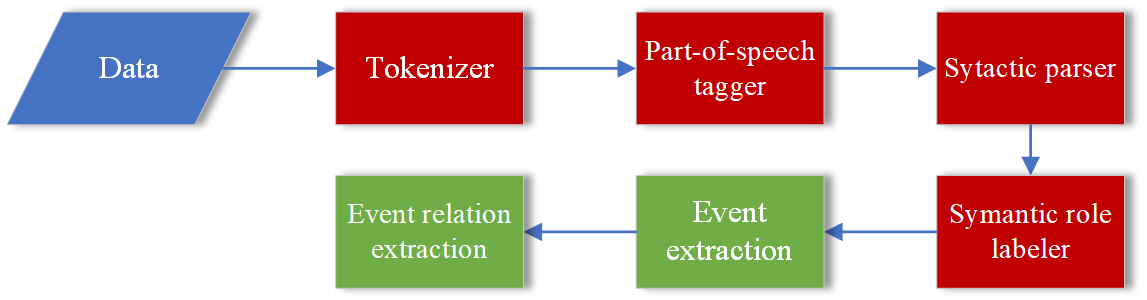}
 \caption{A pipeline model for event extraction and event relation identification}
 \label{fig:pipeline-based}
\end{figure}

Rospocher et al \cite{rospocher2016building} and Li et al \cite{li2017eeg} \cite{li2018constructing} 
have introduced good solutions for these two tasks by using an NLP pipeline. Fig. \ref{fig:pipeline-based} shows the common components in their pipeline-based approach. 
This approach is flexible because each component can be implemented and replaced independently. However, the disadvantage is that it heavily relies on complicated feature engineering approaches and supervised NLP toolkit, which might lead to error propagation\cite{yu2017machine}. 

\begin{figure}
\centering
  \includegraphics[width=.9\linewidth]{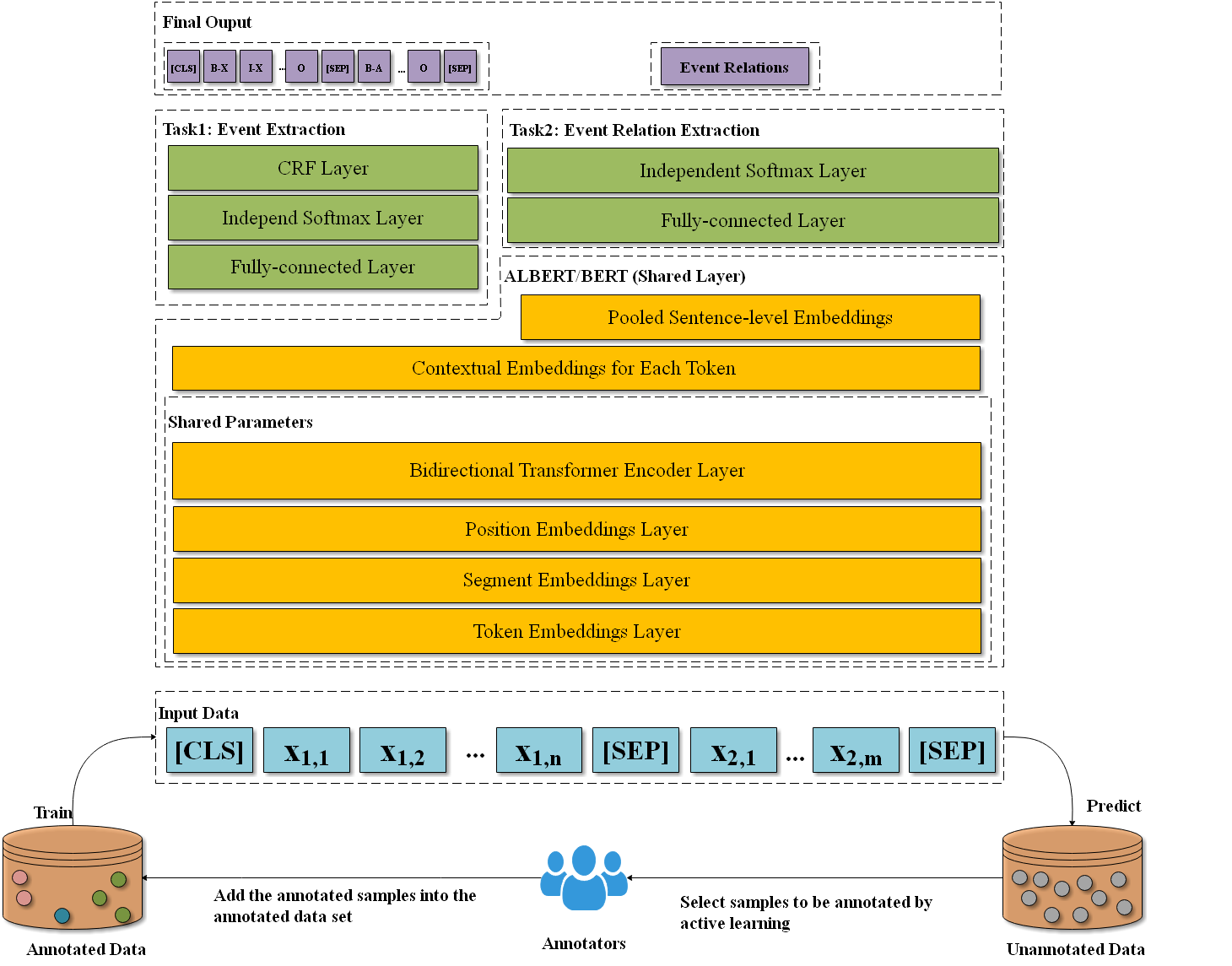}
 \caption{A joint learning end-to-end model and active learning process}
 \label{fig:e2emodel}
\end{figure}

To address this disadvantage, we transform the \textit{pipeline-based} model into an \textit{end-to-end} model, which is shown in Fig. \ref{fig:e2emodel}.
Joint learning is used to deal with the relevance of two target tasks that are identified as green components in Fig. \ref{fig:pipeline-based} to improve the performance based on a shared layer that uses a pretrained language model to learn the potential links between them. Considering that all the data-driven approaches require rich annotated data, but in practice, there is a lack of enough annotated datasets for the two tasks, we introduce active learning in the model to efficiently build high-quality service event datasets.

\subsubsection{Data Representation} 
\begin{figure*}[htbp]
 \centering
 \includegraphics[width=0.9\textwidth]{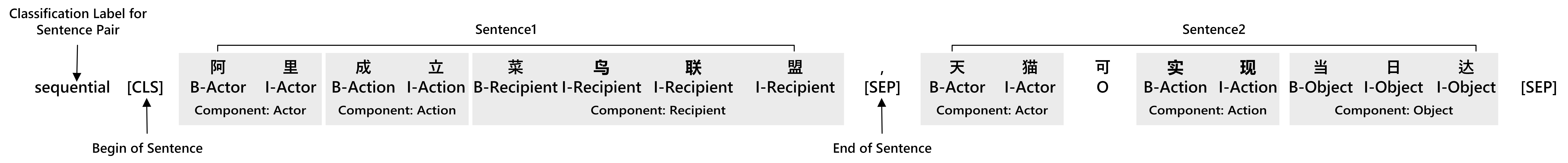}
 \caption{An example of an annotated sample ($[PAD]$ tags are not shown). This news was publicized on March 28, 2016. }
 \label{fig:labeled_sample}
\end{figure*}

For the input news titles, two adjacent sentences are combined as a pair. In particular, if a news title has only one sentence, a blank sentence is added to form a pair. Each sentence pair is taken as a sample. Each token in a sentence is marked with BIO scheme tags (\textit{begin}, \textit{inside}, and \textit{outside})\cite{ramshaw1999text}, special $[CLS]$ and $[SEP]$ tokens are added to distinguish the boundary of the sentence, and $[PAD]$ tokens are added at the end of the tag sequences to make the length of all tag sequences uniform. These token-level tags are mainly used for extracting event components. The sentence pair itself also has a classification label to indicate the relations between two events. Fig. \ref{fig:labeled_sample} gives an example of the annotated data.

Formally, each sentence $\mathbf{x}$ in length $n$ is denoted as $\mathbf{x} = <x_1, x_2, \dots , x_n>$, and the corresponding tag sequence is denoted as $\mathbf{y} = <y_1, y_2, \dots, y_n>$. $c$ is used to represent the relation between events. Therefore, each training sample is denoted as a quintuple $(c, \mathbf{x}_1, \mathbf{x}_2, \mathbf{y}_1, \mathbf{y}_2)$.

\subsubsection{Shared Layer} 
The shared layer is designed to learn the potential links between the two tasks mentioned above by sharing network layer parameters. We use a pretrained language model (such as BERT and ALBERT) as our shared layer, which encodes each token $x_i$ into a fixed-length vector $v_i$ and encodes each sentence $\mathbf{x}$ into a fixed-length
vector $\mathbf{v}$. In our experiment, both the token-level embedding vector and sentence-level embedding vector have $d$ dimensions.

\subsubsection{Event Extraction Module}
The input of this module is a token-level embedding sequence $v = <v_1, v_2, \dots, v_n>$, which is the output of the shared layer. This module consists of a fully connected
layer with \textit{softmax} and a linear-CRF layer spliced behind. For a token $v_i$ at position $i$, the mathematical expression for the fully connected layer is as follows:

\begin{equation}\label{equ:fullyconnect}
 z_i = W^{T}v_i + b
\end{equation}
where $W$ is the weight matrix, $b$ is bias, 
and $K$ is the number of tag categories.

\textit{Softmax} is used to calculate the probability of each tag category:

\begin{equation}\label{equ:softmax}
 h_i = \frac{e^{z_i}}{\sum_{j=1}^K e^{z_{i,j}}}
\end{equation}

We use $h_{i,j}$ to denote the probability that the token at position $i$ gets the tag $j$ and use $h = <h_1, h_2, \dots, h_n>$ to represent a sequence of probabilities of different token tags.

A linear-chain CRF module defines $\mathbf{y}$'s posterior probability, given $\mathbf{x}$:

\begin{equation} \label{equ:crf}
    P(\mathbf{y}|\mathbf{x};A) = 
        \frac{1}{Z(\mathbf{x})}\exp \left(h_{1, y_1} + \sum_{j=1}^{n-1} h_{j+1, y_{j+1}} + A_{y_j,y_{j+1}} \right)
\end{equation}
where $Z(\mathbf{x})$ is a normalization factor over all possible tags of $\mathbf{x}$, and $y_j$ is the tag at position $j$. $A$ is a parameter called the transfer matrix, which can be set manually or by module learning. $A_{p, q}$ is the probability of a transition from tag $p$ to $q$. 
$\mathbf{y}^{*}$ is used to represent the most likely tag sequence of $\mathbf{x}$, which is also the output of this module to represent the event component sequence.

\begin{equation}
    \mathbf{y}^{*} = \arg\max_{\mathbf{y}} P(\mathbf{y}|\mathbf{x}; A)
\end{equation}

The parameter $A$ is learned through the maximum log-likelihood estimation, that is, to maximize the log-likelihood function $\ell_{1}$ of training set sequences in the annotated data set $\mathcal{L}$:

\begin{equation}
    \ell_{1}(\mathcal{L};A) =  \sum_{l=1}^{L}\log P(\mathbf{y^{(l)}}|\mathbf{x^{(l)}};A) 
\end{equation}
where $L$ is the size of the tagged set $\mathcal{L}$. $\ell_{1}$ is a loss function for the event extraction module.

\subsubsection{Event Relation Identification Module}

This module takes two sentence-level embeddings
$\mathbf{v_1}$ and $\mathbf{v_2}$ as its input. It consists of a fully connected layer and a \textit{softmax} layer, and their formulations are similar to Eq. (\ref{equ:fullyconnect}) and Eq. (\ref{equ:softmax}).

The output of this module is $\hat{c}=<\hat{c}_1, \hat{c}_2, \dots, \hat{c}_M>$, where $M$ is the number of predefined types of relations, $\hat{c}_i$ is the probability of the  $i$-th event relation category and there is
\begin{equation}
 \sum_{i=1}^{M} \hat{c}_i = 1
\end{equation}

We use the cross-entropy loss function for this module. $\ell_2$ can be calculated as follows:

\begin{equation}
 \ell_2 = - \sum_{l=1}^{L}\sum_{i=1}^{M} \delta_(c^{(l)},\hat{c}^{(l)}_i)\log(\hat{c}^{(l)}_i)
\end{equation}
where $c$ is the right category, $\delta(i, j)$ is an indicator variable, and if $i=j$, $\delta(i, j)=1$; otherwise, $\delta(i, j)=0$.

The category that has the maximum probability is
selected as the output of the layer, which represents the relation between two events.

It is important to note that there are only \textit{sequential} relations between events in MSEM. If a \textit{reverse sequential} relation between two events $(e_1, e_2)$ is identified, the sequence of the two events should be swapped, and a \textit{sequential} relation from $e_2$ to$e_1$ is added to the MSEM.

\subsubsection{Overall Loss of the Model}
The overall loss of the model can be calculated by:
\begin{equation}
    \ell = \omega_1 \ell_1 + \omega_2\ell_2    
\end{equation}
where $\omega_1$ and $\omega_2$ are used to control which task the model is more biased towards.

\subsubsection{Active Learning}
As mentioned previously, a high-quality annotation dataset must be built from scratch to train the joint learning model. Active learning can help reduce the considerable burden of manual annotation\cite{shen-etal-2017-deep}.

The largest challenge in active learning is how to select instances that need to be 
manually annotated. A selection strategy $\phi (\mathbf{x})$ is a function used to 
evaluate each instance $\mathbf{x}$ in the unlabeled pool $\mathcal{U}$ and to select the 
most informative instances $\{\mathbf{x}\}$ for manual annotation.

\begin{algorithm}[htbp]
\caption{Pool-based active learning framework}
\label{alg:pool_based}
\begin{algorithmic} 
\REQUIRE Labeled dataset $\mathcal{L}$, \\
        \quad \quad \quad Unlabeled data pool $\mathcal{U}$, \\
        \quad \quad \quad Selection strategy $\phi(\cdot)$, \\
        \quad \quad \quad Query batch size $B$
\WHILE{\textbf{not} reach stop condition}
\STATE $//$ Train the model using labeled set $\mathcal{L}$
\STATE $train(\mathcal{L})$;
\FOR{$b=1$ to $B$}
\STATE $//$select the most informative instance
\STATE $\mathbf{x}^{*} = \arg\max_{\mathbf{x}\in\mathcal{U}}\phi(\mathbf{x})$
\STATE $\mathcal{L} = \mathcal{L} \cup <\mathbf{x}^{*}, label(\mathbf{x}^{*})>$
\STATE $\mathcal{U} = \mathcal{U} - \mathbf{x}^{*}$
\ENDFOR
\ENDWHILE
\end{algorithmic}
\end{algorithm}

Algorithm \ref{alg:pool_based} illustrates the pool-based active learning process. In general, samples may expose more rich information in complex tasks, which means it is easier to select samples with rich information through complex tasks. The event extraction task is more complex than the event relation identification task;
therefore, the sample selection is based on the event extraction module. In the following, all active learning strategies mentioned are applied to the event extraction module.

We propose a novel selection strategy called the \textbf{lowest token probability (LTP)}, which selects the token with the lowest probability under the most likely tag sequence $\mathbf{y}^{*}$.

\begin{equation}\label{eq:ltp}
    \phi^{LTP}(\mathbf{x}) = 1 - \min_{y_i^* \in \mathbf{y}^*}h_{i, y^{*}_i}
\end{equation}

Different from traditional selection strategies such as \textbf{minimum token probability (MTP)} and \textbf{least confidence (LC)} \cite{culotta2005reducing}, which only consider local information (i.e., probability $h$ of each token) or global information (i.e., the confidence of the whole sentence sequence $p(\mathbf{y}*|\mathbf{x})$), our \textbf{LTP} selection strategy considers both global and local information; thus, it can select more informative samples and reduce annotation cost.

\subsection{Phase 3: Model Fusion and Evolutionary Part Construction}\label{sec:phase3}

\subsubsection{Model Fusion}\label{sec:fusion}

This step (S5 in Fig. \ref{fig:construction}) aims to fuse the results of Phase 1 (structural part of stakeholder layer, service \& feature layer, and domain layer) and Phase 2 (event layer) by connecting event components with entities in the structural part. In other words, it identifies the cross-layer structural relations in Fig. \ref{fig:overview} such as \textbf{HasActor}, \textbf{HasRecipient}, and \textbf{HasObject}. Challenges for this task are as follows:

\begin{itemize}

\item Entities in event components are extracted from a public news corpus, and due to the openness of natural language representation, it is very common that multiple mentions correspond to the same entity in the structural part of MSEM. For example, both \textit{Ali} and \textit{Alibaba} refer to \textit{Alibaba Group Holding LTD}, which is a formal and full name appearing in KGs.

\item Because news has higher timeliness in reporting the latest events occurring in the real world while KGs usually have longer lags, some mentions in event components cannot be linked to specific entities in the structural part of the MSEM. For example, \textit{Tencent Meeting}, which was released on December 25, 2019, was not included in the generic KG \textit{ownthink} until February 2020.

\end{itemize}

For an entity mentioned in event components, if there is the same entity in the structural part of MSEM, then they are linked directly. If there is no such direct-mapping entity in MSEM, the API\footnote{\url{https://api.ownthink.com/kg/knowledge?entity=entity_name}}
provided by \textbf{ownthink} is used to find the corresponding entity. Because our SKG is extracted from the generic KG provided by \textbf{ownthink}, this API works well for this purpose. If the API returns null, then a new entity corresponding to the mention in event components is created and added into the structural part of the MSEM.

\subsubsection{Evolutionary Relation Generation}
The final step (S6 in Fig. \ref{fig:construction}) is to generate evolutionary relations between elements in the \textit{stakeholder layer} and between elements of the \textit{stakeholder layer} and \textit{service \& feature layer} by making use of the dynamic information enclosed in events. This is the core step of the construction process.

We use a rule-based approach. Each rule is manually defined by experts and describes a specific mapping pattern from an event to one or multiple evolutionary relations. A rule is composed of a set of trigger words, a sequence of event components and a set of evolutionary relations to be constructed:

\begin{equation}
    rule = [Twords, <c_1, ..., c_n>, Edges]
\end{equation}
where $Twords$ is a set of trigger words, and if they appear in the event texts, this rule is triggered; $<c_1, ..., c_n>$ is a sequence of event components; and $Edges$ is a set of evolutionary relations. Each $(c_{i1}, ..., c_{m5} ) \in Edges$ represents an evolutionary relation that has five components (\textit{source}, \textit{destination}, \textit{relation}, \textit{timestamp}, \textit{additionalAttributes}), and $c_{ij}(j\le 5)$ is mapped to a specific event component $c_i (1 \le i \le n)$ in $<c_1, ..., c_n>$ or remains null.

In our practice, a rule base is first prepared based on our perception of massive service-related events we have identified. Now, there are 60 rules in total, and readers may refer to GitHub\footnote{\url{https://dwz.cn/udPC4pFE}} for details of these rules. For an incoming event to be processed, each rule in the rule base is taken out to match the event; if a matching rule is found, an evolutionary relation is constructed in terms of the rule and added into MSEM. Note that since model fusion has been accomplished in Section \ref{sec:fusion}, the evolutionary relation can be directly added into the stakeholder layer of the  MSEM.

Those events that cannot find any corresponding matching rules are clustered by means of text clustering, and then the categories with a large number of samples are selected and interpreted by human experts to explain the commonality of these events and define a new rule for these events. What needs to be explained is that manual intervention is necessary because rules for evolutionary relation construction rely heavily on the long-term accumulation of human knowledge in the service domain.

\section{Quantitative Comparison with Other Service Ecosystem Models}\label{sec:compare}
\begin{table*} 
\centering
\caption{Quantitative comparison between MSEM and existing service-related ecosystem models}\label{tab:compare}
\begin{tabular}{llllll}
\hline
                                                                                                                & Cost          & Coverage                                                                                         & Interpretability & \begin{tabular}[c]{@{}l@{}}Semantic\\ integrity\end{tabular} & \begin{tabular}[c]{@{}l@{}}Semantic\\ accuracy\end{tabular} \\ \hline
                                                                                                              MSEM (Ours)                                                                                                             & \underline{*} & \begin{tabular}[c]{@{}l@{}}\underline{***}\\ (Domain, Business, Technical, Feature)\end{tabular} & \underline{***} & \underline{***}                                              & **                                                          \\ \hline
Writtern et al\cite{wittern2014graph}                                                                                 & ***           & \begin{tabular}[c]{@{}l@{}}**\\ (Technical, Feature)\end{tabular}                                & **              & **                                              & \underline{***}                                             \\ \hline
\begin{tabular}[c]{@{}l@{}}Huang et al\cite{huang2014recommendation}\\ Han et al\cite{adeleye2018constructing}\end{tabular} & **            & \begin{tabular}[c]{@{}l@{}}*\\ (Technical)\end{tabular}                                          & *               & *                                                            & *                                                          \\ \hline
Wang et al\cite{wang2018dkem}                                                                                         & ***           & \begin{tabular}[c]{@{}l@{}}*\\ (Technical)\end{tabular}                                          & **              & *                                                            & \underline{***}                                             \\ \hline
Zhang et al\cite{zhang2015dynamic}                                                                                    & **            & \begin{tabular}[c]{@{}l@{}}*\\ (Business)\end{tabular}                                           & *               & **                                                           & *                                                          \\ \hline
\end{tabular}
\end{table*}

In this section, we quantitatively compare our service ecosystem model MSEM with several existing service-related artificial ecosystem models and demonstrate the advantages of the MSEM.

Artificial ecosystem model representations can be classified into six categories \cite{franco2017open}: tabular representations, meta models, class diagrams, ad hoc notations, conceptual maps, and social networks. The first five categories are static and cannot support dynamic evolution analysis, and only social network-based approaches are data-driven and can handle dynamic information. Our MSEM falls into a social network-based category.

Table \ref{tab:compare} shows the comparative results of several social network-based service ecosystem models from the following aspects:
\begin{itemize}
    \item \textbf{Cost}: cost of constructing a large-scale model for a real-world service ecosystem; the cost mainly comes from the difficulty in obtaining enough external data and extracting the necessary information from the data.
\item \textbf{Coverage}: the number of different semantics layers a service ecosystem model can cover.
\item \textbf{Interpretability}: ability of a model to demonstrate the \textit{evolution} of the service ecosystem and to explain the \textit{causes} of evolution.
\item \textbf{Semantic integrity}: ability of a service ecosystem model to delineate all kinds of real interconnections between services without ignoring valuable ones.
\item \textbf{Semantic accuracy}: accuracy of the description of interconnections between services.
\end{itemize}

\textbf{Cost:}
\cite{wittern2014graph} and \cite{wang2018dkem} constructed network models by monitoring the running state of services. Practically, it is difficult to detect this state of external services that are out of the scope of the organization due to the commercial and legal barriers, etc. \cite{huang2014recommendation} and \cite{adeleye2018constructing} used the services/APIs information that are publicized by service providers; however, the number of services that have enough exposed information is very limited and only contains 16,518 API nodes (including \textbf{1,525} nonisolated nodes only), and the exposure of such information usually has a higher time lag. The MSEM approach could actively perceive changes in a service ecosystem through public news media, which makes it more feasible for collecting massive data in a timely and less expensive manner. The model we constructed in our experiment contains more than \textbf{140,000} nonisolated nodes. This indicates that our model is approximately \textbf{100} times larger, and more importantly, our model continues to grow along with the collection of more news corpora.

\textbf{Coverage}: the MSEM covers multiple layers of service ecosystems in the real world, including interconnections between domains, business-level services, technical-level services, and the most fine-grained functional and nonfunctional features. Models in \cite{huang2014recommendation, adeleye2018constructing, wang2018dkem} focused on the technical aspect (APIs, API mashups, and microservices). Models in \cite{zhang2015dynamic} focused on the business aspect (companies). Models in \cite{wittern2014graph} covered the technical aspect (APIs and mashups) and the feature aspect (characteristics of APIs and mashups). The MSEM model is compatible with these models. For example, the API service ecosystem model is a subgraph of \textbf{channel} interconnections in the \textbf{stakeholder layer}.

\textbf{Interpretability:} The introduction of events enables our model to explain the causes of evolutionary relations and further explain the driving forces of local or global changes in a service ecosystem. This is because journalists manually ''perceive'' such changes and news reported by them contain rich information about the changes. However, traditional models do not have such a corresponding mechanism: they can only model the results of service ecosystem changes but cannot determine why such changes occur, especially from a business perspective. Compared with \cite{adeleye2018constructing, huang2014recommendation, zhang2015dynamic}, models in \cite{wittern2014graph, wang2018dkem} are in control of a run-time system, and thus, execution logs can be easily obtained to help explain possible reasons for the technical evolution of the system.

\textbf{Semantic integrity:} Traditional models contain a set of predefined relation types, such as \textit{invoking} relations between MSs\cite{wang2018dkem} or mashup services and APIs \cite{huang2014recommendation, adeleye2018constructing}; \cite{zhang2015dynamic} includes \textit{cooperation} and \textit{competition} relations between companies; and \cite{wittern2014graph} contains \textit{invocation} relations among users, APIs and applications, as well as \textit{ownership} relations between APIs and characteristics. Different from these models, the MSEM allows us to extend more types of semantically rich evolutionary relations, which covers detailed interconnection semantics more comprehensively and thus has higher semantic integrity. Compared with \cite{huang2014recommendation,adeleye2018constructing, wang2018dkem}, the models in \cite{wittern2014graph, zhang2015dynamic} contain more types of relations with additional information. The relation in \cite{wittern2014graph} is supplemented with information such as invoke time and invoke status, while the relation in \cite{zhang2015dynamic} is supplemented with scores to indicate the degree of competition and cooperation. In summary, in terms of semantic integrity, the models in \cite{wittern2014graph, zhang2015dynamic} are better than the models in \cite{huang2014recommendation,adeleye2018constructing, wang2018dkem}.

\textbf{Semantics accuracy:} Because approaches in\cite{wang2018dkem, wittern2014graph} monitor run-time service states within the scope of a closed service system, their models contain the most accurate description of interconnections between service entities. Approaches in\cite{huang2014recommendation, adeleye2018constructing} convert a bipartite graph (mashup-APIs) into an API social network graph based on an assumption that there should be a connection between two different APIs that are called by the same mashup. Unfortunately, this assumption is not always true and may also involve many interconnections that should not exist. In models of \cite{zhang2015dynamic}, there is at most one relation between any two companies, either \textit{competition} or \textit{cooperation}. In practice, the MSEM model extracts various event information from unstructured news text, which can represent multiple relations between any two organizations/services. Although the limitation of NLP techniques might introduce some inaccurate relations, most of the multirelations identified in the MESM model are reasonable in reality.

In summary, our MSEM approach outperforms state-of-the-art service-related ecosystem models in terms of \textbf{cost}, \textbf{coverage}, \textbf{interpretability}, and \textbf{semantic integrity}; for \textbf{semantic accuracy}, our MSEM approach is not the best, but there is still space for improvement. To the best of our knowledge, the MSEM is by far the most suitable model for service ecosystem modeling.

\section{Experiments and Evaluation}\label{sec:eval}
\subsection{Datasets}\label{sec:dataset}

To build the SKG, we choose \textit{ownthink} as the generic knowledge graph for the following reasons:
\begin{itemize}
    \item \textbf{Easy access:} \textit{ownthink} is an open source Chinese knowledge graph project that provides many APIs for querying and operating its knowledge graph. These APIs help us reduce the workload.
    \item \textbf{Large scale:} \textit{ownthink} is the largest Chinese knowledge graph with the data in CSV format and a total of \textbf{140 million} triples.
\end{itemize}

To let our SKG cover more service-related entities and have more detailed information, we collect extra \textit{organizations} and \textit{executives} from \textit{PEdaily}, which contains the above $86,500$ companies. We also collect additional \textit{organizations}, \textit{channels}, \textit{services} and \textit{features} from \textit{MI App store}, which contains more than $20,000$ popular apps. We also obtain information from other data sources, and readers can obtain details of these data on GitHub.

To extract high-quality service-related events from the news corpora, authoritative and reliable news websites that are focused on modern Internet-based services are carefully selected, including \textit{36Kr} and \textit{tmtpost}.
A complete list of data sources used in this experiment can be found on GitHub. In total, \textbf{358,374} news titles that occurred between December 25, 2004, and December 8, 2019, were collected.
Considering the increase in social media in recent years, most of the events occurred after 2014.

\subsection{Performance of MSEM Structural Part Construction}\label{sec: structural_construction}

Following the steps mentioned in Section \ref{sec:knowledge}, we built an SKG\footnote{
https://dwz.cn/NIMfk8ZH} containing \textbf{116,757} nodes and \textbf{210,761} links. Table 
\ref{tab:knowledge_sta} shows statistics of the SKG.

\begin{table}[htbp]
\caption{Statistics of the SKG}\label{tab:knowledge_sta}
\begin{tabular}{@{}cllcll@{}}
\toprule
\multicolumn{3}{c}{\#Stakeholders} & \multirow{2}{*}{\begin{tabular}[c]{@{}c@{}}\#Services  \\ \& \\Features\end{tabular}} & \multirow{2}{*}{\#Domains} & \multirow{2}{*}{\#Links} \\ \cmidrule(r){1-3}
Executives & \multicolumn{1}{c}{Orgs} & \multicolumn{1}{c}{Channels} &  &  &  \\ \midrule
\multicolumn{1}{l}{2,816} & 55,095 & 24,346 & {32,389} & 2,111 & 210,761 \\ \bottomrule
\end{tabular}
\end{table}

The main cost of constructing the SKG comes from manually filtering service-unrelated entities from \textit{ownthink}. This task was performed by three volunteers.
and approximately 90 hours were used in total. 
A number of service-related entities (including well-known organizations, expired channels, obscure services, and so on) are randomly selected to test the quality of the SKG, and over $85\%$ of these entities can be found in the SKG. This shows that the SKG has high quality and can support building a high-quality MSEM.

\subsection{Performance of Event Extraction and Event Relation Identification}
\subsubsection{Parameter Settings}

In this section, we give the detailed parameters of the joint learning model and the active learning algorithm introduced in Section \ref{sec:extraction_news}.

For the joint learning model, we use
\textbf{ALBERT}\cite{lan2019albert}
as our pretrained language model, which has $4~M$ parameters in total.
The training batch size is set to $32$, and the $max\_seq\_length$ is set to 128. The learning rate is set to $0.00002$. In total, $40$ epochs are trained for convergence.

Other parameters related to ALBERT are set to default values. In the fully connected layer, \textbf{dropout} is set to $0.25$ to prevent overfitting. In the settings of the
parameter transfer matrix $A$, we first give it an initial value that does not require prior knowledge of data distribution, as long as a large penalty is given to the impossible transfer sequence (such as from \textbf{B-Actor} to \textbf{I-Action}); then, the model learns the parameter by itself. Task bias weights $\omega_1$ and $\omega_2$ are both set to 1. For the event relation extraction task, we set additional auxiliary categories including \textit{sequential}, \textit{unrelated} and \textit{reverse sequential} mentioned in Section \ref{sec:extraction_news}, and the case where the sample contains only a single sentence, and the two sentences in the sample together represent an event. Thus, the total relation category $M$ is $5$.

For the active learning selection strategy, the only parameter that needs to be set is the query batch size $B$. To balance the model training time and manual annotation time, we set the value of $B$ to $50$.

\subsubsection{LTP Performance Evaluation}
We evaluate LTP on both the benchmarks and real-world data. The detailed performance on benchmarks can be found in \cite{liu2020ltp}. For real-world data,
we define the cost of each sample annotation as follows:

\begin{equation}
 cost = |T_p \cup T_r| - |T_p \cap T_r|
\end{equation}
where $T$ is the set of tags, and each element in $T$ can be represented as a triple $<start, end, tag>$,
Specifically, if $start=end=1$, then this triple denotes an event relation. $T_p$
denotes the tag set given by model prediction, and $T_r$ is the right tag set. This cost
represents the number of manual operations required for a sample.
\begin{figure}[hbtp]
 \includegraphics[width=\linewidth]{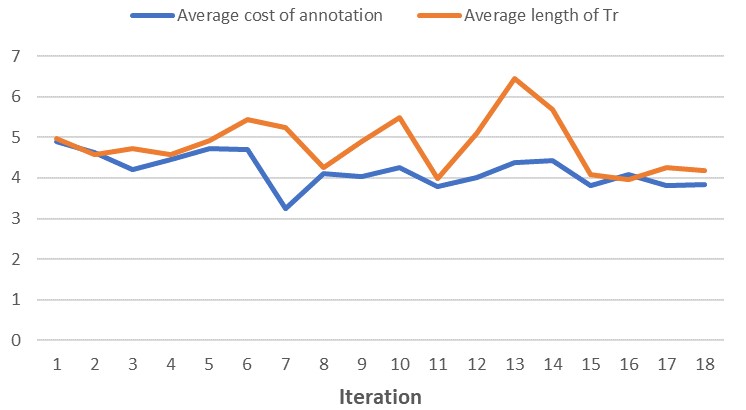}
 \caption{Average annotation cost and average $T_r$ length of each iteration}
 \label{fig:anotation_cost}
\end{figure}

Fig. \ref{fig:anotation_cost} illustrates the average annotation cost and average $T_r$ length of each iteration.
The blue line shows that the average annotation cost of each sample is between $3$ and $5$, and the
annotation cost is slowly decreased. The orange line is located above the blue line after 2 iterations, which
indicates that the use of model preannotation can help reduce manual annotation cost. The gap between
the two lines shows an expanding trend, which also indicates that the model improves the accuracy of
event component (NOT event) extraction and event relation extraction.

\subsubsection{Overall Performance Evaluation on Event Extraction and Event Relation Identification}\label{sec:qe}
In this section, we show the quality evaluation of event extraction and event relation identification by our joint
learning model that is trained using data sets selected by $17$ rounds of active learning ($1050$ annotated samples).
As there lacks a standard for comparison, human judgment is adopted. For event extraction,
if all the event components are correctly extracted, then we consider the event to be correctly
identified. For event relation identification, if the predicted relation is the same as the actual relation,
we consider the event relation to be correctly identified.

We randomly selected $samples$ from the dataset and divided them into $4$ equal parts, each containing $100$
samples (labeled $D_1$, $D_2$, $D_3$, $D_4$). Each part was submitted to a pair of human raters,
which independently evaluated each sample of their part. The average score of two raters is the final result of this part.

\begin{table}[hbtp]
\centering
\caption{Quality evaluation of joint learning model}
\label{tab:eva_jlm}
\begin{tabular}[c]{llllll}
\hline
 & $D_1$ & $D_2$ & $D_3$ & $D_4$ & \textbf{Overall}\\ \hline
\begin{tabular}[c]{@{}l@{}}Event extraction\\ accuracy\end{tabular} & 0.74 & 0.66 & 0.68 & 0.79 & \textbf{0.718}\\
\begin{tabular}[c]{@{}l@{}}Event relation\\ identification accuracy\end{tabular} & 0.84 & 0.90 & 0.88 & 0.92 & \textbf{0.885} \\ \hline
\end{tabular}
\end{table}

Table \ref{tab:eva_jlm} presents the resulting accuracy on the whole evaluation dataset, as well as the accuracy 
on each part. The results show that for the event extraction task, the overall accuracy is $0.718$, and the accuracy 
ranges from $0.66$ to $0.79$ on each part. For the event relation identification task, the overall accuracy is $0.885$, and 
the accuracy ranges from $0.84$ to $0.92$ on each part. Such accuracy is acceptable, and along with the accumulation of more annotated data, the accuracy can be improved continuously.

\subsection{Performance of Model Fusion and Evolutionary Relation Generation}

\subsubsection{Performance of Model Fusion}
For entity linking, we randomly selected 100 entities for evaluation. $27$ entities were found to have no direct correspondence in our knowledge graph, and only two entities were incorrectly mapped in the results using the \textit{ownthink} API. This kind of error mapping mainly comes from entity ambiguity, which will be considered in future work.

\subsubsection{Performance of Evolutionary Relation Generation}
At the time of writing, we summarized $60$ evolutionary relation generation rules, and approximately $46.5\%$ of the samples can be explained by these rules. We used the evaluation method in Section \ref{sec:qe} to evaluate the evolutionary relation triples generated from samples meeting the rules, with an accuracy of $95\%$. It is important to note that coverage will increase as the number of rules we summarize grows.
After applying the evolutionary relation generation rules on the extracted events, we extracted $93,812$ stakeholders, $100,969$ service \& features, and $283,172$ evolutionary relations from $358,374$ news titles.

\section{Applications Scenarios of MSEM} \label{sec:application}
In this section, we give two application scenarios of MSEM, especially from the perspective of service ecosystem evolution analysis, to demonstrate the usability of MSEM.

\subsection{Evolution Analysis on a Stakeholder in terms of its Features}
\begin{figure*}
    \centering
    \includegraphics[width=0.9\textwidth]{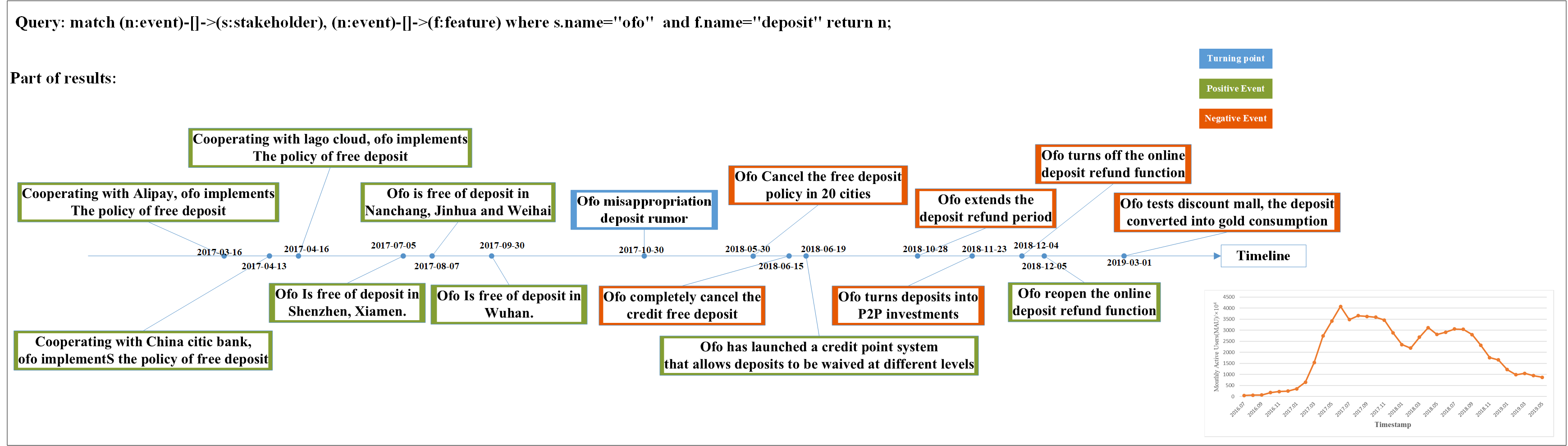}
    \caption{The evolution of a bike-sharing company \textit{ofo} in terms of its deposit service and the MAU of \textit{ofo}}
    \label{fig:ServiceEvolutionExample}
\end{figure*}

Exploring the evolution of a single stakeholder in terms of one of its features is of great significance for discovering the evolutionary roadmap, which consists of key actions on the interconnections with other stakeholders on this feature, i.e., what decisions this stakeholder has made in terms of this feature. The result of these decisions/actions is demonstrated by the changes in the position that the stakeholder holds in the service ecosystem and can be used for reference by other stakeholders.

Our MSEM has the capacity to construct the evolution storyline for one stakeholder and one of its features by using a simple query to retrieve relevant events from the model. Fig. \ref{fig:ServiceEvolutionExample} shows the evolution of a bike-sharing company \textit{ofo} in
terms of a feature \textit{deposit-free} through a cipher query statement. From this storyline, users can easily discover the following facts:

\begin{itemize}
 \item In the growth stage (from February 2017 to May 2017), \textit{ofo} cooperated with 
 other stakeholders to promote the \textit{deposit-free} service feature.
 \item \textit{ofo}'s massive, independent promotion of its \textit{deposit-free} service marks the \textit{heyday} of its deposit-free bike riding service (from July 2017 to October 2017).
 \item The event \textit{"ofo misappropriated the deposit"} can be seen as a \textit{turning point} of \textit{ofo}'s deposit-free service.
 \item From May 2018 to July 2018, \textit{ofo} gradually canceled the deposit-free service, indicating that the service began to wither. It attempted to take advantage of deposit waivers to keep things from getting worse.
 \item The situation worsened after October 2018: \textit{ofo} has great difficulty in refunding user deposits.
\end{itemize}

The rationality of the above analysis can be demonstrated by the number of monthly active users (MAU) of \textit{ofo}, as shown in the lower right corner of Fig. \ref{fig:ServiceEvolutionExample}. It is clear that the MAU curve is closely correlated to the evolution of events.

\subsection{Evolution Analysis of a Service Ecosystem}

As mentioned in Section \ref{sec:introduction}, \textbf{how} and \textbf{why} a service
ecosystem evolution are two key objectives of service ecosystem evolution analysis. Making use of evolutionary relations extracted from events, our MSEM model has the capacity to interpret the evolution phenomena and the corresponding driving forces.

The analysis process can be mapped to a \textit{community evolution tracking} problem \cite{dakiche2019tracking}
in the field of social network analysis, which consists of the following steps:

\begin{itemize}
    \item MSEMs of a given service ecosystem at different times are constructed following the approach in Section \ref{sec:construction}, i.e., to recover a set of snapshots of the service ecosystem.
    \item Static community detection algorithms are applied to these snapshots to obtain the service community structure in each service ecosystem snapshot.
    \item For two adjacent MSEM snapshots, their community structures are aligned, i.e., to identify identical communities at different times.
    \item The change degree of identical communities is measured, and a set of \textit{evolution events} is identified. An evolution event represents a drastic evolution of the service ecosystem, including \textit{birth}, \textit{death}, \textit{split}, \textit{merge} of a community. These evolutionary events are used to show how the service ecosystem evolves.
    \item A set of original service events that cause the appearance of each evolution event is identified from the model, and then high-level semantic and sequential pattern analysis is carried out on these events to summarize the driving force of each evolution event from the business perspective.
\end{itemize}

\begin{figure}
    \centering
    \includegraphics[width=0.9\linewidth]{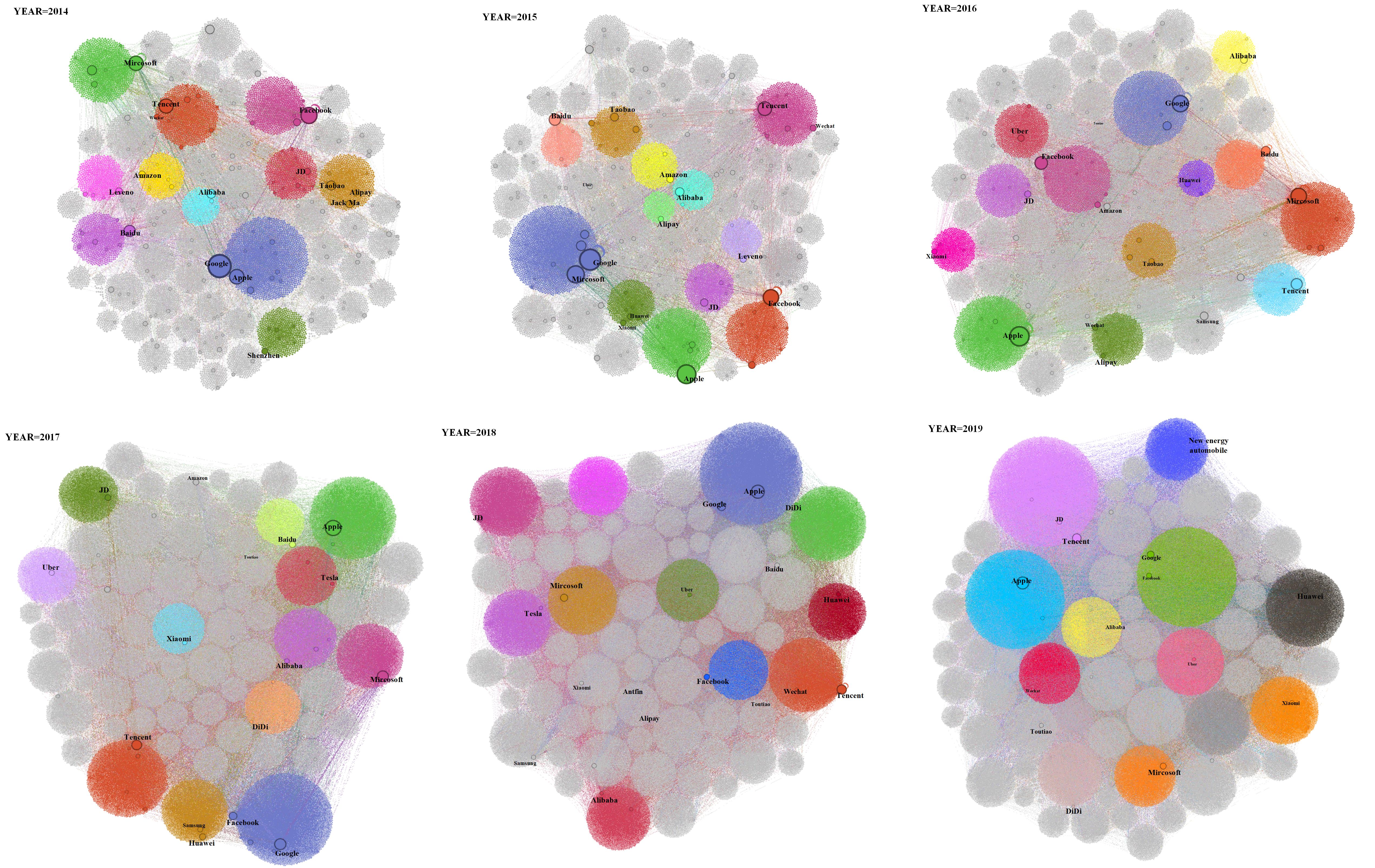}
    \caption{A visual way to track service community evolution.}
    \label{fig:community_evolution}
\end{figure}

Fig. \ref{fig:community_evolution}\footnote{Higher resolution images: \url{https://bit.ly/393X2SF}} shows the evolution process of a service ecosystem visually. 
Each circle in the diagrams is a community, and those communities having large sizes are colored. It is easy to see the core stakeholders those giant communities such as \textit{Google}, \textit{Tencent} and \textit{Alibaba}.

We use key nodes for alignment between communities at different times and find that some communities remain stable (\textit{Tencent}, \textit{Alibaba}), some are shrinking (\textit{Baidu}, \textit{Amazon}), and some are expanding (\textit{Huawei}, \textit{Toutiao}). Through cluster analysis on the events that lead to the evolution of a \textit{Huawei}-centered community, we find that a core factor that drives the expansion of this community in 2019 is the flourish of 5G-related investment, research and development.

It should be noted that there are still many challenges for traditional community detection algorithms to be directly applied in service ecosystem evolution analysis. For example, in Fig. \ref{fig:community_evolution}
Some communities are detected because stakeholders offer the same functional features, but there are not enough interconnections among them. We are now working on a novel community detection algorithm and the corresponding service ecosystem evolution analysis methodology. The objective is to identify rational evolution phenomena and the corresponding driving forces to offer significant insights to stakeholders to help them make 
decisions. Due to limited space, details of the evolution analysis approach cannot be introduced here.
\section{Conclusions and Future Works}
This paper models the Internet of Services (IoS) as a multilayer semantic network from the viewpoint of ecosystems. The significant advantages of MSEM are twofold:

\begin{enumerate}

    \item It covers not only high-level domains and middle-level business services but also fine-grained technological service functional and nonfunctional features so that the characteristics of a service ecosystem can be fully exhibited and explored.

    \item By incorporating ``service events'' into the model, we give MSEM the ability to exhibit the continuous evolution of a service ecosystem. Events can be obtained from public news with low cost and can be utilized for analyzing the drivers of evolution. This empowers the MSEM’s more explanatory power, i.e., it is more interpretable.
\end{enumerate}

In addition to MSEM specifications, a data-driven approach for MSEM construction is introduced. This method overcomes the shortcomings of traditional methods in building large-scale service ecosystems in two ways: (1) Numerous news corpora are continuously collected, and service events are extracted from these massive unstructured texts so that rich real-world data can be used; (2) high-quality open source KGs and external data sources are also utilized to enrich MSEM with more information.

Finally, although two real-world examples of how to make use of the model for service ecosystem evolution analysis are included in this paper, there is still a lack of details. Our future work will provide a solid and systematic method for exploring the evolutionary patterns of service ecosystems and identifying the intrinsic drivers of evolution patterns. Our ultimate goal is to empower service providers to obtain accurate and timely insights into service innovation opportunities in the global service market.

\section*{Acknowledgment}
Research in this paper is partially supported by the National Key Research and Development 
Program of China (No 2018YFB1402500), the National Science Foundation of China 
(61832004, 61772155, 61802089, 61832014).

\ifCLASSOPTIONcaptionsoff
  \newpage
\fi

\bibliographystyle{IEEEtran}
\bibliography{reference}

\begin{IEEEbiography}[{\includegraphics[width=1in,height=1.25in,clip,keepaspectratio]{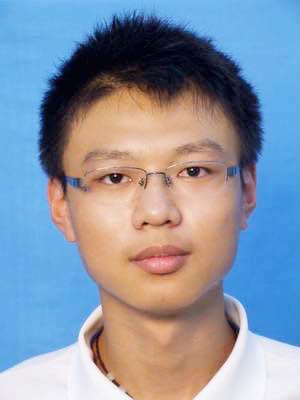}}]{Mingyi Liu}
received his B.S. degree from the School of Computer Science and Technology, Harbin Institute of Technology in 2018. He is currently pursuing the Ph.D. degree in software engineering at Harbin Institute of Technology (HIT), China. His research interests include service ecosystem model, service evolution analysis, data mining and knowledge graph.
\end{IEEEbiography}
\vspace{-10 mm}
\begin{IEEEbiography}[{\includegraphics[width=1in,height=1.25in,clip,keepaspectratio]{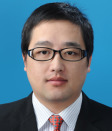}}]{Zhiying Tu}
 is a associate profess of School of Computer Science and Technology at Harbin Institute of Technology (HIT). He holds a PhD degree in Computer Integrated Manufacturing (Productique) from the University of Bordeaux. Since 2013, He began to work at HIT. His research interest is Service Computing, Enterprise Interoperability, and Cognitive Computing. He has 20 publications as edited books and proceedings, refereed book chapters, and refereed technical papers in journals and conferences. He is the member of IEEE Computer Society, and CCF China.
\end{IEEEbiography}
\vspace{-10 mm}
\begin{IEEEbiography}[{\includegraphics[width=1in,height=1.25in,clip,keepaspectratio]{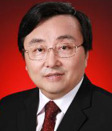}}]{Xiaofei Xu}
is a profess or at School of Computer Science and Technology, and vice president of Harbin Institute of Technology. He received the Ph.D. degree in computer science from Harbin Institute of Technology in 1988. His research interests include enterprise intelligent computing, services computing, Internet of services, and data mining. He is the associate chair of IFIP TC5 WG5.8, chair of INTEROP-VLab China Pole, fellow of China Computer Federation (CCF), and the vice director of the technical committee of service computing of CCF. He is the author of more than 300 publications. He is member of the IEEE and ACM.
\end{IEEEbiography}
\vspace{-10 mm}
\begin{IEEEbiography}[{\includegraphics[width=1in,height=1.25in,clip,keepaspectratio]{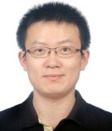}}]{Zhongjie Wang}
 is a profess or at School of Computer Science and Technology, Harbin Institute of Technology (HIT). He received the Ph.D. degree in computer science from Harbin Institute of Technology in 2006. His research interests include services computing, mobile and social networking services, and software architecture. He is the author of more than 40 publications. He is a member of the IEEE.
\end{IEEEbiography}

\end{document}